
\input phyzzx
%
%
\newcount\lemnumber   \lemnumber=0
\newcount\thnumber   \thnumber=0
\newcount\conumber   \conumber=0

\def\myeq{{\rm \chapterlabel\the\equanumber}}

\def\Lemma{\par\noindent\global\advance\lemnumber by 1
           {\bf Lemma\ (\chapterlabel\the\lemnumber)}}
\def\Corollary{\par\noindent\global\advance\conumber by 1
           {\bf Corollary\ (\chapterlabel\the\conumber)}}
\def\Theorem{\par\noindent\global\advance\thnumber by 1
           {\bf Theorem\ (\chapterlabel\the\thnumber)}}

%
%
\def\e{\adveq\eqno{\rm (\chapterlabel\the\equanumber)}}
\def\adveq{\global\advance\equanumber by 1}
\def\twoline#1#2{\displaylines{\qquad#1\hfill(\adveq\myeq)\cr\hfill#2
\qquad\cr}}

\def\manyeq#1{\eqalign{#1}\e}

%
%
\font\tensl=cmsl10
\font\tenss=cmssq8 scaled\magstep1
\outer\def\quote{
   \begingroup\bigskip\vfill
   \def\endquote{\endgroup\eject}
    \def\par{\ifhmode\/\endgraf\fi}\obeylines
    \tenrm \let\tt=\twelvett
    \baselineskip=10pt \interlinepenalty=1000
    \leftskip=0pt plus 60pc minus \parindent \parfillskip=0pt
     \let\rm=\tenss \let\sl=\tensl \everypar{\sl}}
\def\from#1(#2){\smallskip\noindent\rm--- #1\unskip\enspace(#2)\bigskip}

\def\CALT{\address{Division of Physics, Mathematics
and Astronomy\break
Mail Code 452--48\break
California Institute of Technology\break
Pasadena, CA 91125}}

\def\r#1{$\lb \rm#1 \rb$}

%
%
\def\rarrow{\rightarrow}
\def\d#1{{\rm d}#1\,}

\def\semidirect{\mathrel{\raise0.04cm\hbox{${\scriptscriptstyle |\!}$
\hskip-0.175cm}\times}}

\def\mod{\mathop{\rm mod}\nolimits}

\def\ref#1{$^{#1}$}

\def\wiggle{\tilde}

\def\half{{1\over2}}
\def\lb{\lbrack}
\def\rb{\rbrack}

\def\diam{{\hbox{\hskip-0.02in
\raise-0.126in\hbox{$\displaystyle\bigvee$}\hskip-0.241in
\raise0.099in\hbox{ $\displaystyle{\bigwedge}$}}}}
\def\dop{\mathop{{\diam}}\limits}
\def\dw#1#2#3#4{
{\scriptstyle{#4}}\,
{\dop_{#3}^{#1}}
{\scriptstyle{#2}}  }
\def\bw#1#2#3#4#5{{w\left(\matrix{#1&#2\cr#3&#4\cr}\bigg\vert #5\right)}}
\def\sw#1#2#3#4#5{{S\left(\matrix{#1&#2\cr#3&#4\cr}\bigg\vert #5\right)}}
\overfullrule=0pt
\Pubnum={}
\pubtype={Caltech preprint}
\date{February, 1993}
\titlepage
\title{On RSOS Models Associated to Lie Algebras and RCFT}
\author{Doron Gepner\foot{On leave from: Department of Nuclear Physics,
Weizmann Institute of Science, Rehovot, Israel.}}
\CALT
\abstract
RSOS models based on the Lie algebras $B_m$, $C_m$ and $D_m$
are derived from the braiding of conformal field theory.
This gives the first systematic derivation of these models
earlier described by Jimbo et al. The general two field
Boltzmann weights associated to any RCFT are described, giving
in particular the off critical thermalized Boltzmann weights.
Crossing properties are discussed and are shown to agree with
the general theory which connects these with toroidal modular
transformations. The soliton systems based on these lattice
models are described and are conjectured based on the mass formulae and
the spins of the integrals of motions to describe
perturbations of the RCFT $G_k\times G_1\over G_{k+1}$, where
$G$ is the corresponding Lie algebra.
\endpage
\def\wig{\wiggle}
\par
\REF\Found{D. Gepner, Caltech preprint CALT--68--1825, November 1992}
\REF\GepFuchs{J. Fuchs and D. Gepner, Caltech preprint CALT--68--1843,
January 1993}
\REF\Jimbo{M. Jimbo, T. Miwa and M. Okado, Lett. Math. Phys. 14 (1987)
123}
In ref. \r\Found\ the author has suggested a generic way of constructing
solvable interaction round the face lattice models, and their associated
fusion soliton systems, using directly the results of conformal field theory.
The Boltzmann weights are related to the braiding matrices of the rational
conformal field theory (RCFT). In ref. \r\GepFuchs\ such two block
braiding matrices were computed for the general two block RCFT
using the analytic properties of the correlation functions.
Our aim in this note is to further establish this correspondence by
revisiting the $B_m$, $C_m$ and $D_m$ RSOS models described in ref.
\r\Jimbo, and to rederive them as fusion IRF models based on RCFT,
along with describing the associated soliton systems which solve
perturbations of the coset RCFT ${G_k\times G_1\over
G_{k+1}}$ where $G$ is any of the algebras.

Let us return to the expression for the two block braiding derived
for a general RCFT in ref.
\r\GepFuchs. It was found there that the braiding matrix in
the channel $\langle \phi(z) \phi_s(0)\phi_t(1)\phi_u(\infty)\rangle$ is
given by
$$B=\sigma\pmatrix{d&\rho\cr \rho& -d^{-1} \cr},\e$$
where
$$d_i=e^{-\pi i(\Delta+\Delta_u-2\Delta_i^s)},\qquad
d=\sqrt{d_1/d_2},\e$$
$$\sigma=\sqrt{{\sin(\pi\beta+\pi\gamma)\sin(\pi\beta+\pi\delta)
\over \sin(\pi\alpha)\sin(\pi\beta)
d_1 d_2}},\e$$
$$\rho=\sqrt{-{\sin(\pi\gamma)\sin(\pi\delta)
\over \sin(\pi\beta+\pi\gamma) \sin(\pi\beta+\pi\delta)}},
\e$$
where the exponents $\alpha$, $\beta$, $\gamma$ and $\delta$ are defined
by
$$\alpha=\Delta_2^s-\Delta_1^s+1,\e$$
$$\beta=\Delta_2^t-\Delta_1^t,\e$$
$$\gamma=\Delta_1^s+\Delta_1^t+\Delta_1^u-\Delta-\Delta_s-\Delta_t-\Delta_u,
\e$$
$$\delta=\Delta_1^s+\Delta_1^t+\Delta_2^u-\Delta-\Delta_s-\Delta_t-
\Delta_u,\e$$
and the dimensions are defined by $\Delta=\Delta(\phi)$, $\Delta_i^a=
\Delta(\phi_i^a)$ where $\phi_i^a$ is the field appearing in the $a$
channel, $\phi\cdot \phi_a=\phi_1^a+\phi_2^a$, where $a$ is $s$, $t$ or
$u$.

Now suppose that the fields $\phi_s$ and $\phi_t$ are the same, or
that we are interested in the braiding of two identical fields.
This is the case relevant for the model $IRF({\cal O},x,x)$,
where $x=\phi_s=\phi_t$ in the notation of
ref. \r\Found. Note that all
the four exponents can now be expressed in terms of $\alpha$ and
the crossing parameter $\lambda=\pi (\Delta_2^u-\Delta_1^u)/2$,
$$\beta=\alpha-1,\qquad \delta=1-\alpha+{\lambda\over\pi},\qquad
\gamma=1-\alpha-{\lambda\over\pi}.\e$$
Substituting into eq. (1), the braiding matrix
may be written as
$$B=\hbox{``phase''}\times {1\over \{\alpha\}}\pmatrix{
-e^{-i\pi \alpha} & \sqrt{\{1-\alpha-\wig \lambda\}
\{1-\alpha+\wig\lambda\}}\cr *& e^{i\pi \alpha}\cr},\e$$
and we denoted $\wig\lambda=\lambda/\pi$ and
$\{x\}={\sin(\pi x)\over \sin\lambda}$.

Recall now from ref. \r\Found\ that from the braiding matrix of the RCFT
one builds the Boltzmann weight of the solvable fusion interaction
 round the face (IRF) model IRF$({\cal O},x,x)$ as follows.
 One defines the Hecke algebra element associated to the braiding matrix
 by
 $$H_i=H\pmatrix{\phi& p\cr q& \phi_u\cr}=e^{-i\lambda}-B_{pq},\e$$
 where $B_{pq}$ is the braiding matrix, eq. (10) and
 $p$ and $q$ are the primary fields which label the conformal blocks
 in the $s$ channel. The $H_i$ obey the usual relations for the Hecke
 algebra,
 $$\eqalign{H_i H_{i+1} H_i-H_i&=H_{i+1} H_i H_{i+1} -H_{i+1},\cr
     H_i H_j&=H_j H_i,\qquad{\rm for\  |i-j|\geq2},\cr
     H_i^2&=(2\cos\lambda) H_i,}\e$$
from which it follows that we may build the solvable face transfer
matrix,
$$X_i(u)=\sin(\lambda-u)+\sin u \cdot H_i,\e$$
where $u$ is the spectral parameter. The Hecke algebra implies that
the face transfer matrix $X_i(u)$ obeys the star--triangle equation
(STE),
$$\eqalign{X_i(u) X_{i+1}(u+v) X_i(v)&=X_{i+1}(v) X_i(u+v) X_{i+1}(u),\cr
   X_i(u)X_j(v)&=X_j(v) X_i(u),\cr}\e$$
from which it follows that the transfer matrices for different
values of the spectral parameter $u$ commute, enabling the exact
solution of the model.

Substituting the braiding matrix eq. (10) into the expression for the
face transfer matrices, we find the Boltzmann weights of the
lattice model IRF$({\cal O},x,x)$ which are as follows.
In case there is only one block:
$$\dw \phi p {\phi_u} p=\{\wig\lambda(1-u)\}.\e$$
For the two block case, it is found
$$\manyeq{
\dw \phi p {\phi_u} p&={\{\alpha+\wig\lambda u\} \over \{\alpha\}},\qquad
\dw \phi q {\phi_u} q={\{\alpha-\wig\lambda u\} \over \{\alpha\}},\cr
\dw \phi q {\phi_u} p&=\dw \phi p {\phi_u} q=
\epsilon\{\wig\lambda u\}{\sqrt{
\{\alpha+\wig\lambda\} \{ \alpha-\wig\lambda\} } \over \{\alpha\} },\cr}$$
 where $\epsilon=\pm1$ labels complex conjugate solutions and
 $p=\phi_1^u$ and $q=\phi_2^u$ are the two intermediate fields in the
 $u$ channel.
 It is straightforward to verify that the Boltzmann weights of the
 model IRF$(SU(N),N,N)$ given in
ref. \r{\Jimbo,\Found,\GepFuchs}
agree precisely with the above general formula, eqs. (15-16), when one
substitutes the RCFT $SU(N)_k$.

  Now, it is possible to contemplate the generalization of the above
  Boltzmann weights off criticality. This is done by simply redefining
  the symbol $\{x\}$ to be
  $$\{x\} ={\Theta_1(\pi x,p) \over \Theta_1(\lambda,p) },\e$$
  where the parameter $p$ labels the distance from criticality and the
  theta function is defined by
  $$\Theta_1(u,p)=2p^{1\over 4} \sin u \prod_{n=1}^\infty
  [1-2p^{2n} \cos(2u)+p^{4n} ] (1-p^{2n}).\e$$
  In the critical limit $p\rarrow 0$ we recover the previous definition
  of $\{x\}$ and thus obtain the same Boltzmann weights as before.
  This is the expression for the off critical Boltzmann weights by
  merely substituting the new definition of $\{x\}$ into eqs. (15-16).
  It should be possible to verify that this Boltzmann weights obey
  the STE for all values of $p$ and thus define a thermalized solvable
  lattice model.
  In particular, in the case of $SU(N)$ we recover the
  thermalized Boltzmann weights previously given in ref. \r\Jimbo.
\par
Let us turn now to models associated to the other Lie algebras.
These are the restricted solid on solid (RSOS) lattice models first
described in ref. \r\Jimbo, which are associated with the Lie algebras
$B_m$, $C_m$ and $D_m$ in the usual Cartan notation. We wish to
revisit these models in light of the connection with RCFT put forwards
in ref. \r\Found.
The Boltzmann weights of these RSOS models are \r\Jimbo
$$\dw d {d+\mu} {d+2\mu} {d+\mu}=
{[\lambda-u] [\omega-u] \over [\lambda] [\omega]},\qquad{\rm for\ }
\mu\neq 0,\e$$
$$\dw d {d+\mu} {d+\mu+\nu} {d+\mu}={[\lambda-u] [d_{\mu \nu}+u]
\over [\lambda] [d_{\mu\nu}]},\qquad {\rm for\ } \mu\neq\pm\nu,\e$$
$$\dw d {d+\nu} {d+\mu+\nu} {d+\mu}={[\lambda-u] [u]\over
[\lambda][\omega]} \cdot \left({[d_{\mu\nu}+\omega][d_{\mu\nu}-\omega]
\over [d_{\mu\nu}]^2} \right)^\half,\qquad{\rm for\ }
\mu\neq\pm\nu,\e$$
$$\dw d {d+\nu} d {d+\mu}={[u] [d_{\mu-\nu}+\omega-\lambda+u]\over
[\lambda] [d_{\mu-\nu}+\omega] } (g_{d\mu} g_{d\nu})^\half+
\delta_{\mu\nu} {[\lambda-u] [d_{\mu-\nu}+\omega+u] \over
[\lambda] [d_{\mu-\nu}+\omega]},\qquad {\rm for\ }\mu\neq0,\e$$
$$\dw d d d d={[\lambda+u] [2\lambda-u]\over [\lambda] [2\lambda]}-
{[u] [\lambda-u]\over [\lambda][2\lambda]} J_{d0}.\e$$
Here $d$ stands for an arbitrary integrable highest weight of the respective
algebra, at the level $k$, where $k$ is some integer.
$\mu$ and $\nu$ are arbitrary elements of the set $\Sigma$
which is defined as, $\Sigma=\{0,\pm e_1,\pm e_2,\ldots,\pm e_m\}$
for $B_m$ and $\Sigma=\{ \pm e_1,\pm e_2,\ldots,\pm e_m\}$ for
$C_m$ and $D_m$, and where $e_i$ are orthonormal set of unit vectors in
the canonical basis of the algebras. We have used the symbol
$[x]=\Theta_1(x;p)$ where the theta function was defined in eq. (18).
At criticality, $p=0$, and we find $[x]=\Theta_1(x,0)
\propto \sin(x)$. Here $d_{\mu\nu}$ stands for
$d_{\mu\nu}=\omega(d+\rho,\mu-\nu)$ and $d_\mu=d_{\mu 0}$. Further,
$$g_{d\mu}=\sigma {s(d_\mu+\omega)\over s(d_\mu)}
\prod_{\kappa\neq \pm u,0} {[d_{\mu\kappa}+\omega]\over [d_{\mu\kappa}]
},
\qquad {\rm for\ } \mu\neq 0,\qquad g_{d0}=1.\e$$
$$J_{d0}=\sum_{\kappa\neq0} {[d_k+\half\omega-2\lambda]\over
[d_k+\half\omega]} g_{d\kappa}.\e$$
The parameters are $\sigma=1$, $\lambda=(2m-1)\omega/2$ and $s(z)=[z]$
for the $B_m$ model, $\sigma=-1$, $\lambda=(m+1)\omega/2$ and $s(z)=
[2z]$ for the $C_m$ model and $\sigma=1$, $\lambda=(m-1)\omega$ and
$s(z)=1$ for the $D_m$ model. For all algebras $\lambda=g\omega/2$,
where $g$ is the dual Coxeter number.

Now, note that the RSOS models based on $B_m$,
$C_m$ and $D_m$ may be interpreted
as fusion interaction round the face lattice models.
The only difference between the restricted and unrestricted SOS models
is that
$\omega$ becomes fixed to the value $\omega={\pi \over k+g}$
and the representations that can appear are
only the ones which are integrable representations at the level $k$.
Importantly, the admissibility condition for the models
is simply the fusion rules with respect to the vector representation
whose highest weight is $\lambda=e_1$.
In the notation of ref. \r\Found\
these are the fusion IRF models $IRF(B_m,v,v)$, $IRF(C_m,v,v)$ and
$IRF(D_m,v,v)$. We wish to verify that the Boltzmann weights are indeed
the specialization of the general ones described in ref. \r\Found.

The product of the vector representation with itself contains
three representations,
$$v^2=1+s+a,\e$$
where $1$ is the singlet, $a$ is the anti--symmetric tensor and $s$
is the symmetric tensor.
Let us now compute the crossing parameters which
are given by \r\Found\ $\zeta_i=\pi(\Delta_{i+1}-\Delta_i)/2$ where
$\Delta_i$ is the conformal dimension of the $i$th field in the
operator product expansion of $v$ with itself; arranged in the
order $1$, $\lambda_2$ and $2\lambda_1$ (this is so that the symmetry
of the representation will be alternating). The dimension of a WZ field
with highest weight $\lambda$ is
$$\Delta_\lambda={\lambda (\lambda+2\rho)\over 2(k+g)},\e$$
where $\rho$ is half the sum of positive roots and the value of the dual
Coxeter number is $g=2m-2$ for $D_m$,
$g=2m-1$ for $B_m$ and $g=m+1$ for $C_m$. A straightforward
calculation shows that the two crossing parameters of the models are
$$\lambda={\pi\Delta_{\lambda_2}\over 2}={\pi g/2\over k+g}
,\qquad
\omega={\pi(\Delta_{2\lambda_1}-\Delta_{\lambda_2})\over2}=
{\pi\over k+g},\e$$
and that $\lambda$ indeed has the values described above.

According to the general theory $\lambda$ is the crossing parameter of
the model and the crossing multiplier should be the toroidal $S$
matrix \r\Found. We wish to check that this is so.
Let us consider then the crossing properties of the amplitudes. It is
known that these Boltzmann weights obey the crossing property
\REF\Wadati{M. Wadati, T. Deguchi and Y. Akutsu, Physics Reports 180
(4\&5) (1989) 247, and ref. therein} \r\Wadati,
$$\bw a b c d u= \bw b c d a {\lambda-\mu} \left[ {\psi(a)\psi(c)\over
\psi(b)\psi(d)}\right]^\half,\e$$
where $\lambda$ plays the role of the crossing parameter and
the crossing multiplier
$\psi(a)$ is given by
$$\psi(d)=\prod_{k=1}^m s(d_k) \prod_{1\leq i<j\leq m}
[d_{ij}] [d_{i-j}].\e$$
Now, according to the general theory of fusion IRF models
ref. \r\Found\ the
crossing multiplier should be identical to the torus modular matrix.
We can see that this is indeed the case for all the models by
making use of the quantum Weyl dimension formula (see e.g.
\REF\LAG{L. Alvarez Gaum\'e, C. Gomez and G. Sierra, Nucl. Phys. B330
(1990) 347}\r\LAG)
$${S_{d,0}\over S_{0,0}}=\prod_{\alpha>0} {\sin[\pi(d+\rho,\alpha)
/(k+g)]\over \sin[\pi (\rho,\alpha)/(k+g)]}.\e$$
Remembering that the positive roots of $C_m$ are $e_i\pm e_j$
for $i<j$, and $2e_i$; $e_i\pm e_j$ and $e_i$ for $B_m$ and
$e_i\pm e_j$ for $D_m$, it is concluded
that eq. (30) is precisely the Weyl quantum dimension
formula (up to the denominator, which is an irrelevant constant) and
so it is established that indeed,
$$\psi(a)\propto S_{a,0},\e$$
for all the three algebras.
We also need to identify $\omega={\pi \over k+g}$ which is
precisely what we
found, eq. (28). This is also the value that allows for the restriction
of the SOS models as remarked above
and so we find a complete agreement on this value
from three points of view.
This establishes the correct crossing properties of the Boltzmann
weights of the $B_m$, $C_m$ and $D_m$ IRF models.
\par
We wish to show now that the Boltzmann weights of the RSOS models
eqs. (19-23) indeed agree with the ones derived,
in general, from conformal
field theory in ref. \r\Found. The $BCD$ models are three field
cases. Namely, the primary field $v$ has three fields in its operator
product, $v^2=\phi_0+\phi_1+\phi_2$ which are, as before,
$\phi_0=1$, $\phi_1=\phi_{\lambda_2}$ and $\phi_2=\phi_{2\lambda_1}$.
It follows that the face transfer matrix is given by the expression
\r\Found,
$$X_i(u)=f_0(u) P^0_i+f_1(u) P^1_i+f_2(u) P^2_i,\e$$
where $f_i(u)$ are the three eigenvalues of the face transfer matrix
which are
$$\twoline{
f_0(u)={\sin(\lambda+u)\sin(\omega+u)\over
\sin\lambda\sin\omega},\qquad
f_1(u)={\sin(\lambda-u)\sin(\omega+u)\over
\sin\lambda\sin\omega},}{
f_2(u)={\sin(\lambda-u)\sin(\omega-u)\over
\sin\lambda\sin\omega},}$$
and the projection operators $P^a_i$ are independent of the spectral
parameter $u$ and are defined by
$$P^a_i=\prod_{q\neq a} {(B_i-\lambda_q) \over
(\lambda_a-\lambda_q)},\e$$
where $B_i$ is the conformal braiding matrix whose eigenvalues are
$$\lambda_a=(-1)^a e^{i\pi (\Delta_a-2\Delta_v)}.\e$$
%

Note now that the projection operator
$$P^a\pmatrix{d & d+\mu\cr d+\mu&d+2\mu\cr}=\delta_{a,2} e^{ i\omega}.\e$$
This follows immediately from the fact that the relevant correlation
function $\langle [d] [v] [v] [d+2\mu] \rangle$ has only one
intermediate block in the $s$ channel which is labeled by the field
$[d+\mu]$. It follows that this correlation function is given by
a rational function of $z$ which is determined by the dimension of the
fields, and so the braiding matrix is
$$B=e^{i\pi (\Delta_2-2\Delta_v)}=e^{i\omega},\e$$
and eq. (37) follows.
{}From eq. (33) it is concluded that
the face transfer matrix obeys
$$X\left( \dw d {d+\mu} {d+2\mu} {d+\mu} \right)=
 {\sin(\lambda-u)\sin(\omega-u)\over \sin\lambda\sin\omega},\e$$
which agrees precisely with the Boltzmann
weights
 described above, eq. (19).
\par
Next consider the projection operators appearing in the block
$P^a\left(\dw d \beta {d+\mu+\nu} \alpha\right)$
where $\alpha,\beta=d+\mu,d+\nu$. Evidently, this projection operator
vanishes for $a=0$. Further, the relevant correlation function is a two
blocks case, and thus use can be made of the preceding formulae, eq.
(15-16).
The relevant correlation function is $\langle [d] [v] [v]
[d+\mu+\nu]\rangle$ and it has exactly two blocks in the $s$ channel
labeled by the blocks $p=[d+\mu]$ or $p=[d+\nu]$. We may thus apply
directly the previous formulae eq. (16) and it follows that the
braiding matrix is, indeed, given by eq. (10),
where we substitute $\omega$
instead of $\lambda$ and where the parameter $\alpha$ computed from
eq. (5) assumes the value
$$\alpha=\Delta_{d+\mu}-\Delta_{d+\nu}=d_{\mu\nu},\e$$
where $d_{\mu\nu}$ was defined prior to eq. (24). Substituting the
eigenvalues of this braiding matrix into the expression for the face
transfer matrix eq. (33) we find the same expression as for the general
two block case, but now in the $1-2$ channel,
$$B=\hbox{``phase''}\times {1\over \{d_{\mu\nu}/\pi\}} \pmatrix{-e^{-{i
d_{\mu\nu}}} & \sqrt{\{(d_{\mu\nu}+\omega)/\pi \} \{
(d_{\mu\nu}-\omega)/\pi \} } \cr *& e^{i{d_{\mu\nu}}} }.\e$$
The face transfer matrix thus assumes the form,
$$\dw d {d+\mu} {d+\mu+\nu} {d+\mu}={[\lambda-u] [d_{\mu\nu}-u] \over
  [\lambda][d_{\mu\nu}]},\e$$
  $$\dw d {d+\mu} {\d+\mu+\nu} {d+\nu}={[\lambda-u][u]\over [\lambda]
  [\omega]} \cdot \left( {[d_{\mu\nu}+\omega] [d_{\mu\nu}-\omega]\over
    [d_{\mu\nu}]^2 } \right)^\half,\e$$
    where we have rescaled the spectral parameter $u\rarrow
   {u\over\lambda}$ and used again $[x]= \sin(x)$.

Lo and behold the expression for the face transfer matrix derived from
conformal field theory, eq. (42-43), is identical precisely to the BCD
Boltzmann weights given above, eq. (19-23), thus providing a first
systematic
derivation of these weights previously found in ref. \r\Jimbo.
Finally, the last two Boltzmann weights appearing in eq. (22-23) involve
the correlation function, $\langle d v v d \rangle$. The number of
blocks in this correlation function is the dimension of the set
$\Sigma$, $|\Sigma|$, and is very large. It follows that it is
difficult to calculate directly these Boltzmann weights from RCFT
except for very small ranks, and this calculation will have to await
the development of the proper generalization to any number of blocks
of the two block calculation of ref. \r\GepFuchs. We may infer the
conformal braiding matrix for this case by calculating the extreme
UV limit of the BCD Boltzmann weights, $u\rarrow i\infty$. We find
the result
$$B_{pq}={[\omega]\over [d_{u-v}+\omega]} e^{-i(d_{u-v}+\omega-\lambda)}
\left[ (g_{d\mu}g_{d\nu})^\half-\delta_{\mu\nu}\right],\e$$
and for $p=q=d$,
$$B_{d d}={2[\omega]\over[2\lambda]} e^{i\lambda}.\e$$
Note that the generalization of the two fields face transfer matrix
to arbitrary temperature, eq. (17) is consistent with the thermalized
Boltzmann weights, eq. (19-23), and
indeed we find the same result. This is
an important check that the thermalization of the general fusion IRF
model suggested above is indeed correct and agrees with all the
known examples.

Let us turn now to integrable soliton systems based on these lattice
models. The vacua of the theory are labeled by the primary fields
of the RCFT $G_k$ where $G$ is one of the Lie algebras $B_m$, $C_m$
or $D_m$. The kinks of the theory interpolate between neighboring
vacua $a$ and $b$ provided that the pair $(a,b)$ obey the admissibility
condition which is, in this case, that  $b-a\in \Sigma$ or that
$b$ appears in the operator product of $a$ with the vector
representation. The two particle scattering matrices for the kink $v$
correspond to the
process $(a|v|b)+(b|v|c)\rarrow (a|v|d)+(d|v|c)$, where we labeled by
$(a|v|b)$ the kink $v$ interpolating between the $a$ and $b$ vacua.
As follows from the general theory of ref. \r\Found, the $S$ matrices of
these kink scattering processes are given by
$$\sw a b c d \theta= F(\theta) \left[ {S_{b,0} S_{c,0}\over
S_{a,0} S_{d,0} }\right]^{\theta/2} \bw a b c d {\lambda\theta},\e$$
where $i\pi \theta$ is the relative rapidity of the incoming particles,
$S$ is the torus modular matrix, $w$ is the Boltzmann weight of
the associated RSOS lattice model and $F(\theta)$ is an overall function
to be determined. As discussed in ref. \r\Found\ the $S$ matrix eq. (46)
obeys the factorization equation for integrable soliton systems as is
guaranteed from the fact that the Boltzmann weights obey the STE.
Further, the $S$ matrix will be crossing invariant and unitary provided
that the function $F(\theta)$ obeys the functional equations
$$\eqalign{
F(\theta)&=F(1-\theta)\cr
F(\theta) F(-\theta)&={1\over\rho(\theta)\rho(-\theta)},\cr}\e$$
where the unitary factor $\rho(\theta)$ is
$$\rho(\theta)={\sin[\lambda(1-\theta)] \sin[\omega-\lambda\theta]
\over [\lambda][\omega] }.\e$$
The minimal solution of this set of functional equations is
$$F(\theta)=f_{2\over g} (\theta) f_{1-{2\over g}}(\theta) \times
Z(\theta) Z(1-\theta),\e$$
where
$$f_\alpha(\theta)={\sin(\pi u/2 +\pi\alpha/2) \over \sin(\pi u/2-
\pi\alpha/2)},\e$$
is the Koberle Sweica amplitude. The piece $Z(\theta)Z(1-\theta)$
is a $Z$ factor that does not have any poles or zeros in the
physical sheet.
A minimal solution for it is
$$\twoline{
Z(\theta)=\exp\bigg(2\int_0^\infty {\d x\over x} {\sinh(g\theta x/2)
\over \sinh[(g+k)x]\sinh(gx)} \cdot
}{
\bigg\{ \cosh(kx/2)
\cosh(gx\theta/2)
-\cosh[(g+k-2)x]\cosh[gx(\theta/2-1)]\bigg\}\bigg).}$$
It is noteworthy that the entire solution for $F(\theta)$ is simply
\REF\Vega{H.J. de Vega and V.A. Fateev, Int. Journal Mod. Phys. A6
(1991) 3221}
a folding of the solution for the $SU(N)$ case ref. \r\Vega\ as
$F(\theta)=K(\theta)K(1-\theta)$, where $K(\theta)$ is the $SU(g)_k$
solution \r\Vega.

$F(\theta)$, eq. (49), has a unique simple pole in the physical strip at
$\theta_b={\omega\over\lambda}$ (except for $B_1$ where the pole lies
outside the physical strip and there are no bound states), along with
the crossing channel pole at $1-\theta_b$. The mass of the first bound
state is thus
$$M_b=2 M_v \cos\left({\pi \omega\over 2\lambda}\right).\e$$
Note that $\lambda=g\omega/2$ where $g$ is the dual Coxeter number
for all the algebras and, thus, the mass of the first bound state
can be written as,
$$M_b=2 M_v  \cos({\pi \over g}).\e$$
Crucially, the mass ratio of the first bound state, and indeed
the entire piece of $F(\theta)$ which contains the poles in the physical
sheet, is independent of the level $k$. It follows that for all values
$k$, the masses of the particles in the theory and their integrals
of motions are the same. This holds also for the $SU(N)$ case,
along with the mass ratio eq. (53). We expect this
to be true for all the other Lie algebras (i.e., the simply laced
$E_6$, $E_7$ and $E_8$ along with the non--simply laced $G_2$ and $F_4$).

To compute the particle spectrum and the spins of the
integrals of motion it is thus enough
to inspect the $k=1$ soliton scattering amplitudes. For the simply
laced algebras the scattering amplitudes reduce to the purely
diagonal scattering amplitudes associated with the ADE algebras
(for a review and references, consult, e.g., ref.
\REF\Klassen{T.R. Klassen and E. Melzer, Nucl. Phys. B 338 (1990)
485}\r\Klassen).
The entire $S$ matrix is given in the $D_m$, $k=1$ case
by $S_{vv}(\theta)=f_{2/g}(\theta) f_{1-2/g} (\theta)$, for the
scattering of two vector solitons.

For the non--simply laced algebras (i.e., $B_m$,
$C_m$, $G_2$ and $F_4$) the mass ratios and the integrals of motion
reduce to those of the corresponding classical Toda theory
\REF\Braden{H.W. Braden, E. Corrigan, P.E. Dorey and R. Sasaki,
Nucl. Phys. B338 (1990) 689}
\REF\Deltwo{G.W. Delius,
M.T. Grisaru, S. Penati and D. Zanon, Nucl. Phys. B359 (1991)125}
\REF\Delius{G.W. Delius,
M.T. Grisaru and D. Zanon, Nucl. Phys. B382 (1992) 365}
refs. \r{\Braden,\Deltwo,\Delius}.
Note, that the amplitudes described in these references are
a factor in the amplitudes we find for the RSOS theories, but
that, however, the full amplitude is not purely diagonal even for $k=1$
and is truly an RSOS theory.
Curiously, in the Toda case it is claimed \r{\Deltwo,\Delius} that the
classical results are not valid and need to be renormalized through
perturbation theory. Here this does not occur and the RSOS theories
described here thus give the first realization
of the non--simply laced classical Toda
system results.
The masses and integrals of
motions are the same as those of the classical Toda systems, as well.
This, indeed, agrees with our derived mass
formula eq. (53). The reminder of the soliton
amplitudes may be found by the
bootstarp procedure or alternatively, the fusion of IRF models.

The particles in the theory are in a one to one correspondence with
the nodes of the Dynkin diagram of the respective algebras and are
labeled by the fundamental weights. Each soliton in the theory is
thus labeled by some $\lambda_i$ where $\lambda_i$ is the $i$th
fundamental weight. The values of the integrals of motion $\gamma_a^s$
are in a one to one correspondence with the eigenvectors of the Cartan
matrix, which are labeled by the exponent set of the algebra $s$.
The Perron--Frobenius vector, which is the eigenvector with the maximal
eigenvalue, gives the masses of the corresponding solitons, i.e., the
mass of the $\lambda_a$ soliton is $\gamma^1_a$. The $\lambda_a$
soliton mediates the vacua connected by the fusion with respect to the
$\lambda_a$ representation, i.e., it corresponds to the
solvable lattice model IRF$(G,\lambda_a,\lambda_a)$, and can be found
by the fusion procedure. The vector amplitude described here is
fundamental for the $B_m$ and $C_m$ cases, i.e., all the other solitons
are composite particles of the vector soliton. The masses are
(for $B_m$ and $C_m$),
$$M_a=\sin\left({\pi a\over g}\right),\e$$
where $a$ labels the representations ($a=1$ is the vector, and $a=2$ is
the bound state described above, agreeing with eq. (53)).
In the
case of $D_m$ the fundamental amplitudes are actually the spinor and
anti--spinor representations, where the vector is the bound state of
these. Via a bootstrap of
the vector amplitude one can get all the amplitudes for all the
representations
except for the spinor ones. Unfortunately, to find the spinor amplitudes for
$D_m$ with $k>1$, it behooves us to calculate the Boltzmann weights
of the solvable lattice model IRF$(D_m,x,x)$ where $x$ is either the
spinor or the anti--spinor, which is quite a challenge, as they involve
a large number of blocks.
The masses of the $D_m$ solitons are given by
$$M_s=M_{\bar s}=1,\qquad M_a=2\sin\left({\pi a\over g}\right),\e$$
where we labeled the vector as $a=1$ and the anti--symmetric tensors
which are its composites by $a=2,3,\ldots,m-2$.
This mass formula agrees, again, with eq. (53).

The spins of the integrals of motions in the theory are
given by the exponents of the corresponding Lie algebra modulo
the Coxeter number (or the dual one). In the case of $D_m$ the
spins are $1,3,5,\ldots,2m-1,m \mod 2(m-1)$. For $C_n$ and $B_n$ the
spins are given by all odd integers. These are exactly the same
spins of the integrals of motion which are encountered for the
$W$ invariant coset theories ${G_k\times G_1\over G_{k+1}}$, where
$G$ is the corresponding Lie algebra, as perturbed by the operator
$\Phi^{0,0}_{\rm ad}$ (i.e. a singlet in the upper Lie algebras and
an adjoint in the lower).
We thus conjecture that the
RSOS scattering matrices described in this note correspond to the
soliton spectrum and scattering amplitudes of these perturbed
conformal field theories.

We hope that this work further illuminates the various
connections between solvable lattice models, conformal field theory
and soliton scattering theories.
\ack
The author wishes to thank P.E. Dorey for a discussion and
the theory groups of Saclay and CERN for the hospitality while this
work was completed.
\refout
\bye